\documentclass[11pt, a4paper]{article}
\usepackage{authblk}
\usepackage{amsthm}  
\usepackage{amssymb}
\usepackage{amsmath}
\usepackage{graphicx}
\usepackage{hyperref}  
\usepackage{enumerate}
\usepackage{amssymb,color}

\numberwithin{equation}{section} 

\newtheorem{theoremN}[equation]{Theorem}
\numberwithin{theorem}{section}

\newtheorem{lemmaN}[equation]{Lemma}

\newtheorem{notationN}[equation]{Notation}

\newenvironment{proofN}{\trivlist \item[\hskip \labelsep{\bf Proof.}]}{
$\blacksquare$ \endtrivlist   }

\definecolor{MyGreen}{RGB}{29,162,55}

\begin{document}

\title{\bf 
  New identities for the Shannon-Wiener entropy function
  with applications
}

\author{Aiden A. Bruen
\thanks{The author gratefully acknowledges the financial support of the National Sciences and Engineering Research Council of Canada}
}
\affil{
  Adjunct Research Professor,\\
  Carleton University,\\
  {\tt abruen@math.carleton.ca}
}

\date{}
\maketitle

\begin{abstract}
  We show how the two-variable entropy function 
  $H(p,q)=p\log\frac{1}{p}+q\log\frac{1}{q}$
  can be expressed as a linear combination of
  entropy functions symmetric in $p$ and $q$
  involving quotients of polynomials in $p,q$
  of degree $n$ for any $n\ge 2$.
\end{abstract}



%
%
%
\thispagestyle{empty}
\section{Introduction}
In Renyi's {\em A Diary on Information Theory}, \cite{Renyi},
he points out that the entropy formula, i.e.,
$$H(X)=p_1\log\frac{1}{p_1}+p_2\log\frac{1}{p_2}+\cdots+p_N\log\frac{1}{p_N},$$
where logs are to the base~2,
was arrived at, independently, by
Claude Shannon and Norbert Wiener in 1948.
This famous formula was the revolutionary precursor of the information age.
Renyi goes on to say that,
\begin{itemize}
  \item[]
  this formula had already appeared in the work of
  Boltzmann which is why it is also called the
  Boltzmann-Shannon Formula.
  Boltzmann arrived at this formula in connection
  with a completely different problem.
  Almost half a century before Shannon
  he gave essentially the same formula to describe entropy
  in his investigations of statistical mechanics.
  He showed that if, in a gas
  containing a large number of molecules
  the probabilities of the
  possible states of the individual molecules are
  $p_1,p_2,\ldots,p_N$
  then the entropy
  of the system is
  $H=c(p_1\log\frac{1}{p_1}+p_2\log\frac{1}{p_2}+\cdots+p_N\log\frac{1}{p_N})$
  where $c$ is a constant.
  (In statistical mechanics the natural logarithm is used
  and not base 2 \ldots).
  The entropy of a physical system is the measure of its disorder
  \ldots
\end{itemize}

Since 1948 there have been many advances in information theory
and entropy.  A well-known paper of Dembo, Cover and Thomas
is devoted to inequalities in information theory.
Here, we concentrate on equalities.
We show how a
Shannon function
$H(p,q)$
can be expanded in infinitely many ways in an
infinite series of functions
each of which is a linear combination of
Shannon functions of the type $H\bigl(f(p),g(q)\bigr)$,
where $f,g$ are quotients of polynomials of degree $n$ for any
$n\ge 2$.
Apart from its intrinsic interest,
this new result gives insight into the algorithm 
in Section~\ref{section:ShannonLimitAppnsCrypt}
for constructing a common secret key between two
communicating parties ---
see also \cite{Bennett}, \cite{BruenForcinitoMcQuillan}, \cite{BruenForcinitoWehlau}, \cite{MaurerWolf}.

%
%
%
\section{Extensions of a Binary Symmetric Channel}
Recall that a binary symmetric channel
has input and output symbols drawn from
$\{0,1\}$. 
We say that there is a common probability
$q=1-p$ of any symbol being transmitted incorrectly,
independently for each transmitted symbol,
$0\le p\le 1$.

\begin{figure}[!ht]
  \begin{center}
    \includegraphics[trim={20 355 20 160},clip,scale=.75]{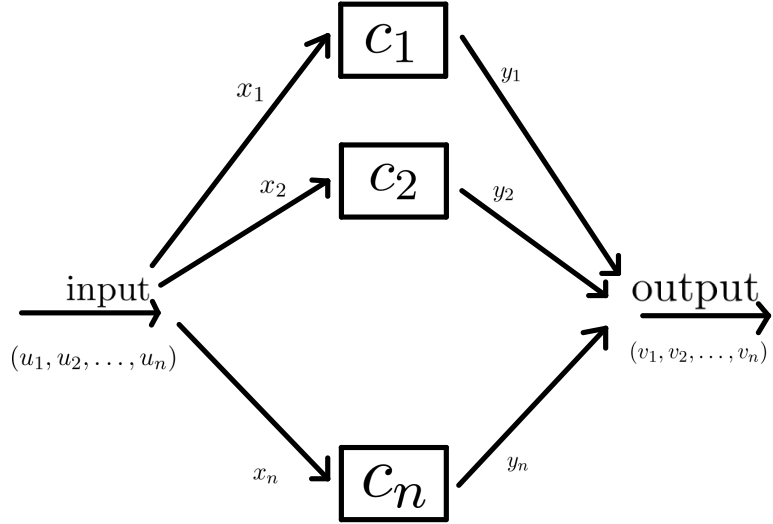}
    \caption{$n$ copies of $C$ acting independently and in parallel.
    \label{figure:nCopiesOfC}}
  \end{center}
\end{figure}

We use the channel matrix
$P=\left(
\begin{array}{cc}
  p&q\\
  q&p
\end{array}\right)$.
Again, $p$ is the {\em probability of success},
i.e., $p$ denotes the probability that 0 is transmitted  to 0
and also the probability that 1 gets transmitted to 1.
The {\em second extension} $P^{(2)}$ of $P$
has  alphabet
$\{00,01,10,11\}$
and channel matrix
$$P^{(2)}=\left(\begin{array}{cccc}
  p^2&pq&qp&q^2\\
  pq&p^2&q^2&qp\\
  qp&q^2&p^2&pq\\
  q^2&qp&pq&p^2
\end{array}\right)
=\left(\begin{array}{cc}
  pP&qP\\
  qP&pP
\end{array}\right).$$
An alternative way to think of an
$n^{th}$ extension of a channel $C$
(see Welsh~\cite{Welsh})
is to regard it as
$n$ copies of $C$ acting independently and in parallel.
See Figure~\ref{figure:nCopiesOfC}.

Let us assume also that, for $C$, the input probability of 0
and the input probability of 1 are both equal to
$\frac{1}{2}$.

\begin{theoremN}
  \label{theorem:XYInputOutput}
  Let $\mathbf{X}=(X_1,\ldots,X_n)$,
  and $\mathbf{Y}=(Y_1,\ldots,Y_n)$
  denote an input-output pair for $C^{(n)}$.
  Then
  \begin{enumerate}[(a)]
    \item
      $H(\mathbf{X})=H(X_1,\ldots,X_n)=n$
    \item
      $H(\mathbf{Y})=H(Y_1,\ldots,Y_n)=n$
    \item
      $H(\mathbf{X}|\mathbf{Y})$ is equal to $nH(p,q)$.
    \item
      The capacity of $C^{(n)}$ is $n\bigl(1-H(p,q)\bigr)$.
  \end{enumerate}
\end{theoremN}
\begin{proofN}
(a)
Since, by definition, the $X_i$ are independent,
$$H(\mathbf{X})=H(X_1)+H(X_2)+\cdots+H(X_n).$$
We have
\begin{align}
H(X_i)&=-\biggl[Pr(X_i=0)\log Pr(X_i=0)+Pr(X_i=1)\log Pr(X_i=1)\biggr]\nonumber\\
&=-\biggl[\frac{1}{2}\log\bigl(\frac{1}{2}\bigr)+\frac{1}{2}\log\bigl(\frac{1}{2}\bigr)\biggr]\nonumber\\
&=-\log\bigl(\frac{1}{2}\bigr)\nonumber\\
&=-[\log 1-\log 2]=\log 2=1.\nonumber
\end{align}
Thus $H(\mathbf{X})=n$.

\noindent
(b) The value of $Y_i$ only depends on $X_i$.
The $X_i$ are independent.
Thus $Y_1,Y_2,\ldots,Y_n$ are independent.
For any $Y_i$ we have
$$Pr(Y_i=0)=Pr(X_i=0)(p)+Pr(X_i=1)(q)=\frac{1}{2}(p+q)=\frac{1}{2}.$$
Also $Pr(Y_i=1)=\frac{1}{2}$.
Then, as for $X_i$, $H(Y_i)=1$.
Thus $H(\mathbf{Y})=H(Y_1,Y_2,\ldots,Y_n)=\sum_i H(Y_i)=n$.

\noindent
(c)
We have
$H(\mathbf{X})-H(\mathbf{X}|\mathbf{Y})
=H(\mathbf{Y})-H(\mathbf{Y}|\mathbf{X})$.
Since $H(\mathbf{X})=H(\mathbf{Y})=n$
we have $H(\mathbf{X}|\mathbf{Y})=H(\mathbf{Y}|\mathbf{X})$.
Now
$$H(\mathbf{Y}|\mathbf{X})=\sum_{\mathbf{x}}Pr(\mathbf{x})H(\mathbf{Y}\,|\,\mathbf{X}=\mathbf{x}),$$
where $\mathbf{x}$ denotes a given value of the random vector $\mathbf{X}$.
Since the channel is memoryless,
$$H(\mathbf{Y}\,|\, \mathbf{X}=\mathbf{x})
=\sum_i H(Y_i\,|\, \mathbf{X}=\mathbf{x})
=\sum_i H(Y_i\, |\, X_i=x_i).$$
The last step needs a little work ---
see \cite{JJ} Exercise 4.10 or \cite{McEliece} 
or \cite{BruenForcinitoMcQuillan} or the example below
for details.
Then
\begin{align}
H(\mathbf{Y}\,| \,\mathbf{X})
&=\sum_{\mathbf{x}}Pr(\mathbf{x})\sum_i H(Y_i\,|\, X_i=x_i)\nonumber\\
&=\sum_i\sum_u H(Y_i\,| \, X_i=u)Pr(X_i=u).\nonumber
\end{align}
Thus
$$H(\mathbf{Y}\,|\,\mathbf{X})=\sum_{i=1}^nH(Y_i\,|\, X_i)=nH(p,q)
  =H(\mathbf{X}\,|\,\mathbf{Y}).$$

\medskip\noindent
{\bf Example}\ \ 
  Let $n=2$ and 
  $\left(\begin{array}{c}x_1\\ x_2\end{array}\right)=
  \left(\begin{array}{c}0\\ 1\end{array}\right)$.
  Then
  $\left(\begin{array}{c}y_1\\ y_2\end{array}\right)=
  \left(\begin{array}{c}1\\ 0\end{array}\right)$ or
  $\left(\begin{array}{c}0\\ 1\end{array}\right)$ or
  $\left(\begin{array}{c}1\\ 1\end{array}\right)$ or
  $\left(\begin{array}{c}0\\ 0\end{array}\right)$.
  Using independence we get that
  the coresponding entropy term is
  $-[q^2\log q^2 + p^2\log p^2 + qp\log qp + pq\log pq]$.
  This simplifies to
  $-2[p\log p + q\log q]=2H(p,q)$.
  Note that the probability that
  $\left(\begin{array}{c}x_1\\ x_2\end{array}\right)=
  \left(\begin{array}{c}0\\ 1\end{array}\right)$
  is $\frac{1}{4}$.

\medskip
\noindent
(d)
The capacity of $C$
is the maximum value, over all
inputs, of $H(X)-H(X\,|\, Y)$.
Since $X$ is random, the input probability of a 1 or 0 is 0.5.
This input distribution maximizes $H(X)-H(X\,|\, Y)$ for $C$,
the maximum value being $1-H(p,q)$.
Then the capacity of $C^{(n)}$ is $n(1-H(p,q))$.
It represents the information about 
$\mathbf{X}$ conveyed by $\mathbf{Y}$ or the amount of information
about $\mathbf{Y}$ conveyed by $\mathbf{X}$.
\end{proofN}
%
%
%
\section{An Entropy Equality}
\label{section:basicEntropyEquality}
First we need some additional discussion on entropy.

\bigskip\noindent
{\bf A. Extending a basic result.}

\medskip\noindent
A fundamental result for random variables $X,Y$ is that
$H(X)+H(Y|X)=H(Y)+H(X|Y)$.
A corresponding argument may be used to
establish similar identities involving more than
two random variables.
For example,
\begin{align}
  H(X,Y,Z)
  &=H(X)+H(Y|X)+H(Z\,|\,X,Y)\nonumber\\
  &=H(X,Y)+H(Z\, |\, X,Y).\nonumber
\end{align}
Also
$$H(X,Y,Z)=H(X)+H(Y,Z\, |\, X).$$

\bigskip\noindent
{\bf B. From Random Variables to Random Vectors.}

\medskip\noindent
For any random variable $X$ taking only a finite number
of values with probabilities $p_1,p_2,\ldots,\allowbreak p_n$
such that
$$\sum p_i=1\qquad\hbox{and}\qquad p_i>0\quad(1\le i\le n),$$
we define the
entropy
of $X$ using the Shannon formula
$$H(X)=-\sum_{k=1}^{n}p_k\log p_k=\sum_{k=1}^n p_k\log\frac{1}{p_k}.$$
Analogously, if $\mathbf{X}$
is a {\em random vector} which takes only a finite number
of values
$\mathbf{u_1},\mathbf{u_2},\ldots,\allowbreak\mathbf{u_m}$,
we define its
entropy
by the formula
$$H(\mathbf{X})=-\sum_{k=1}^m Pr(\mathbf{u_k})\log Pr(\mathbf{u_k}).$$
For example, when $\mathbf{X}$ is a
two-dimensional random vector, say
$\mathbf{X}=(U,V)$ with
$p_{ij}=Pr(U=a_i, V=b_j)$, then we can write
$$H(\mathbf{X})=H(U,V)=-\sum_{i,j}p_{ij}\log p_{ij}.$$
Note that $\sum_{i,j} p_{ij}=1$.

More generally, if $X_1,X_2,\ldots,X_m$
is a collection of random variables each taking
only a finite set of values,
then we can regard
$\mathbf{X}=(X_1,X_2,\ldots,X_m)$
as a random vector
taking a finite set of values and we define the
{\bf joint entropy}
of $X_1,\ldots,X_m$ by
{\small
\begin{align}
&H(X_1,X_2,\ldots,X_m)\nonumber\\
&=H(\mathbf{X})\nonumber\\
&=-\sum Pr(X_1=x_1,X_2=x_2,\ldots,X_m=x_m)\log Pr(X_1=x_1,X_2=x_2,\ldots,X_m=x_m).
\nonumber
\end{align}
}
Standard results for random variables then carry over to
random vectors ---
see \cite{Ash}, \cite{Welsh}.  

\bigskip\noindent
{\bf C. The Grouping Axiom for Entropy.}

\medskip\noindent
This axiom
or identity can shorten calculations.
It reads as follows
(\cite[p. 2]{Welsh}, \cite[p. 8]{Ash},
\cite[Section 9.6]{BruenForcinitoMcQuillan}).

Let $p=p_1+p_2+\cdots+p_m$
and $q=q_1+q_2+\cdots+q_n$
where each $p_i$ and $q_j$ is non-negative.
Assume that $p,q$ are positive with $p+q=1$.
Then
\begin{align}
&
H(p_1,p_2,\ldots,p_m,q_1,q_2,\ldots,q_n)\nonumber\\
&\ \ 
  =H(p,q)+pH\left(\frac{p_1}{p},\frac{p_2}{p},\ldots,\frac{p_m}{p}\right)
  +qH\left(\frac{q_1}{q},\frac{q_2}{q},\ldots,\frac{q_n}{q}\right).
  \nonumber
\end{align}
For example, suppose $m=1$ so $p_1=p$.
Then we get
$$H(p_1,q_1,q_2,\ldots,q_n)=H(p,q)
  +qH\left(\frac{q_1}{q},\ldots,\frac{q_n}{q}\right).$$
This is because
$p_1H\bigl(\frac{p_1}{p_1},0\bigr)=p_1H(1,0)=p_1(1\log 1)=p_1(0)=0$.

\bigskip
\begin{theoremN}
  \label{theorem:z;xy}
  Let $\mathbf{X},\mathbf{Y},\mathbf{Z}$ be random vectors
  such that
  $H(\mathbf{Z}\, |\, \mathbf{X},\mathbf{Y})=0$.
  Then
  \begin{enumerate}[(a)]
    \item
      $H(\mathbf{X}|\mathbf{Y})=H(\mathbf{X},\mathbf{Z}\, |\, \mathbf{Y})$.
    \item
      $H(\mathbf{X}|\mathbf{Y})=H(\mathbf{X}\,|\, \mathbf{Y},\mathbf{Z})+
        H(\mathbf{Y}\, |\, \mathbf{Z})$.
  \end{enumerate}
\end{theoremN}
\begin{proofN}
  \begin{align}
    H(\mathbf{X}\, |\, \mathbf{Y})
    &=H(\mathbf{X},\mathbf{Y})-H(\mathbf{Y})\nonumber\\
    &=H(\mathbf{X},\mathbf{Y},\mathbf{Z})-H(\mathbf{Z}\, |\, \mathbf{X},\mathbf{Y})-H(\mathbf{Y})\nonumber\\
    &=H(\mathbf{X},\mathbf{Y},\mathbf{Z})-H(\mathbf{Y})
      \qquad\hbox{[\,since $H(\mathbf{Z}\,|\,\mathbf{X},\mathbf{Y})=0$\,]}
      \nonumber\\
    &=H(\mathbf{Y})+H(\mathbf{X},\mathbf{Z}\, |\, \mathbf{Y})-H(\mathbf{Y})\nonumber\\
    &=H(\mathbf{X},\mathbf{Z}\, |\, \mathbf{Y}),\qquad\hbox{proving (a).}\nonumber
  \end{align}

  For (b),
  \begin{align}
    H(\mathbf{X}\, |\, \mathbf{Y}) 
    &=H(\mathbf{X},\mathbf{Z},\mathbf{Y})-H(\mathbf{Y})\hbox{ from (a)}\nonumber\\
    &=H(\mathbf{X},\mathbf{Z},\mathbf{Y})-H(\mathbf{Y},\mathbf{Z})
      +H(\mathbf{Y},\mathbf{Z})-H(\mathbf{Y})\nonumber\\
    &=H(\mathbf{X}\, |\, \mathbf{Y},\mathbf{Z})+H(\mathbf{Z}\, |\, \mathbf{Y}).\nonumber
  \end{align}
\end{proofN}

%
%
%
\section{The New Identities}
\label{section:NewIdentities}
We will use the above identity,
i.e.,
\begin{equation}
  \label{equation:BasicIdentity}
  H(\mathbf{X}\, |\, \mathbf{Y})
    =H(\mathbf{X}\, |\, \mathbf{Y},\mathbf{Z})
    +H(\mathbf{Z}\, |\, \mathbf{Y})
\end{equation}
which holds under the assumption that
$H(\mathbf{Z}\, |\, \mathbf{X},\mathbf{Y})=0$.
We begin with arrays
$A=\left(\begin{array}{c}
  a_1\\ a_2\\ \vdots \\ a_n
\end{array}\right)$,
$B=\left(\begin{array}{c}
  b_1\\ b_2\\ \vdots \\ b_n
\end{array}\right)$,
where $n$ is even.
We assume that $A,B$ are random binary strings
subject to the condition that, for each $i$,
we have
$Pr(a_i=b_i)=p$.
We also assume that the events
$\{(a_i=b_i)\}$
form an independent set.
We divide $A,B$ into blocks of size~2.

To start,
put $\mathbf{X}=\left(\begin{array}{c}x_1\\x_2\end{array}\right)$,
$\mathbf{Y}=\left(\begin{array}{c}y_1\\y_2\end{array}\right)$,
$\mathbf{Z}=x_1+x_2$.

\begin{lemmaN}
  $$H(\mathbf{Z}\,|\,\mathbf{X},\mathbf{Y})=0.$$
  \label{lemma:HZXY}
\end{lemmaN}
\begin{proofN}
  We want to calculate
  $\sum_{\mathbf{x},\mathbf{y}}H(\mathbf{Z}\,|\, \mathbf{x},\mathbf{y})Pr(X=\mathbf{x},Y=\mathbf{y})$.
  Given $\mathbf{x},\mathbf{y}$, say
  $\mathbf{x}=\left(\begin{array}{c}\alpha_1\\ \alpha_2\end{array}\right)$,
  $\mathbf{y}=\left(\begin{array}{c}\beta_1\\ \beta_2\end{array}\right)$
  the value of $\mathbf{Z}$ is $\alpha_1+\alpha_2$.
  There is no uncertainty in the value of $\mathbf{Z}$
  given $\mathbf{x},\mathbf{y}$,
  i.e., each term in the above sum for $H$ is $H(1,0)=0$.
  Therefore $H(\mathbf{Z}\,|\,\mathbf{X},\mathbf{Y})=0$.
\end{proofN}

  From this we can use
  formula~(\ref{equation:BasicIdentity})
  for this block of size two.
  We can think of 
  a channel
  from $\mathbf{X}$ to $\mathbf{Y}$ (or from $\mathbf{Y}$ to $\mathbf{X}$)
  which is the second extension of a 
  binary symmetric channel where $p$ is the
  probability of success.
  We have
  $$H(\mathbf{X}\, |\, \mathbf{Y})
    =H(\mathbf{X}\, |\, \mathbf{Y},\mathbf{Z})
    +H(\mathbf{Z}\, |\, \mathbf{Y}).$$
  From Theorem~\ref{theorem:XYInputOutput} part (c) 
  the left side, i.e., $H(\mathbf{X}\,|\,\mathbf{Y})$ is equal to
  $2H(p,q)$.
  Next we calculate the right side
  beginning with
  $H(\mathbf{Z}\,|\,\mathbf{Y})$,
  i.e., $H\left(\mathbf{Z}\,\Big|\,\left(\begin{array}{c}y_1\\y_2\end{array}\right)\right)$.
  We have
  \begin{align}
  H(\mathbf{Z}\,|\,\mathbf{Y})
  &=H\bigl(\mathbf{Z}\,|\,(y_1+y_2=x_1+x_2)\bigr)Pr(y_1+y_2=x_1+x_2)\nonumber\\
  &\qquad +H\bigl(\mathbf{Z}\,|\,(y_1+y_2\ne x_1+x_2)\bigr)Pr(y_1+y_2\ne x_1+x_2).
  \nonumber
  \end{align}
  We know that
  $Pr(x_1+x_2=y_1+y_2)=p^2+q^2$
  and  $Pr(x_1+x_2\ne y_1+y_2)=1-(p^2+q^2)=2pq$
  since $p+q=1$.
  From the standard formula we have
  $H(\mathbf{Z}\,|\,\mathbf{Y})=(p^2+q^2)\log\left(\frac{1}{p^2+q^2}\right)+2pq\log\left(\frac{1}{2pq}\right)$
  since
 $H(\mathbf{Z}\,|\,\mathbf{Y})=H(p^2+q^2,2pq)$.

  Next we calculate
  $$H\left(\left(\begin{array}{cc}x_1\\x_2\end{array}\right)\,\Big |\,
    \left(\begin{array}{c}y_1\\y_2\end{array}\right), (x_1+x_2)\right)
  =H(\mathbf{X}\mid \mathbf{Y},\mathbf{Z}).$$
  Again we have two possibilities, i.e.,
  $y_1+y_2=x_1+x_2$ and $y_1+y_2\ne x_1+x_2$.
  The corresponding probabilities are
  $p^2+q^2$ and $2pq$ respectively.
  We obtain
  $$H(\mathbf{X}|\mathbf{Y},\mathbf{Z})
  =(p^2+q^2)H\left(\frac{p^2}{p^2+q^2},\frac{q^2}{p^2+q^2}\right)
    +2pqH\left(\frac{pq}{2pq},\frac{pq}{2pq}\right).$$
  This comes about from the facts that
  \begin{enumerate}[(a)]
    \item
    If $y_1+y_2=x_1+x_2$ then we either have
    $y_1=x_1$ and $y_2=x_2$ or
    $y_1=1-x_1$, $y_2=1-x_2$.
    \item
    If $y_1+y_2\ne x_1+x_2$
    then either $y_1=x_1$ and $y_2\ne x_2$
    or $y_1\ne x_1$ and $y_2=x_2$.
    \item
    $H(\frac{1}{2},\frac{1}{2})=1$.
  \end{enumerate}
Then from
equation~(\ref{equation:BasicIdentity})
we have
{\em our first identity} as follows
\begin{equation}
  \label{equation:FirstNewIdentity}
  2H(p,q)=(p^2+q^2)H\left(\frac{p^2}{p^2+q^2},\frac{q^2}{p^2+q^2}\right)
    +2pq
    +H(p^2+q^2,2pq).
\end{equation}

\bigskip\noindent
{\bf Blocks of Size Three.}

\medskip\noindent
Here
$\mathbf{X}=\left(\begin{array}{c}x_1\\x_2\\x_3\end{array}\right)$,
$\mathbf{Y}=\left(\begin{array}{c}y_1\\y_2\\y_3\end{array}\right)$,
$\mathbf{Z}=x_1+x_2+x_3$.
As in Lemma~\ref{lemma:HZXY}
we have $H(\mathbf{Z}\,|\,\mathbf{X},\mathbf{Y})=0$
so we can use formula~(\ref{equation:BasicIdentity}) again,
i.e.,
$$H(\mathbf{X}\,|\,\mathbf{Y})=H(\mathbf{X}\,|\,\mathbf{Y},\mathbf{Z})
  +H(\mathbf{Z}\,|\,\mathbf{Y}).$$
We have a
channel from $\mathbf{X}$ to $\mathbf{Y}$
(or from $\mathbf{Y}$ to $\mathbf{X}$)
which is the third extension $C^{(3)}$
of a binary symmetric channel $C$,
where $p$ is the probability that 0 (or 1) is transmitted
to itself.

From Theorem~\ref{theorem:XYInputOutput} we have
$H(\mathbf{X}|\mathbf{Y})=3H(p,q)$.

Similar to the case of blocks of size~2, we have
$H(\mathbf{Z}|\mathbf{Y})=H(p^3+3pq^2,q^3+3qp^2)$.
This is because the probabilities that
$Z=y_1+y_2+y_3$ or
$Z\ne y_1+y_2+y_3$ are,
respectively, $p^3+3pq^2$ or
$q^3+3qp^2$, as follows.

If $Z=y_1+y_2+y_3$, then either
$x_1=y_1$, $x_2=y_2$, $x_3=y_3$
or else, for some $i$, $1\le i\le 3$ (3 possibilities)
$x_i=y_i$ and, for the other two indices $j,k$,
$x_j\ne y_j$ and $x_k\ne y_k$.

A similar analysis can be carried out for the case
where $Z\ne y_1+y_2+y_3$.
We then get
$H(\mathbf{X}\,|\,\mathbf{Y},\mathbf{Z})=f(p,q)+f(q,p)$
where
$$f(p,q)=(p^3+3pq^2)\left\{H\left(\frac{p^3}{p^3+3pq^2},\frac{pq^2}{p^3+3pq^2},
  \frac{pq^2}{p^3+3pq^2},\frac{pq^2}{p^3+3pq^2}\right)\right\}.$$
We now use the grouping axiom for $m=1$.
The $p$ in the formula refers to $\frac{p^3}{p^3+3pq^2}$ here
and the $q$ there is now replaced by
$\frac{3pq^2}{p^3+3pq^2}$.
Then
\begin{align}
  f(p,q)
  &=(p^3+3pq^2)\left\{
  H\left(\frac{p^3}{p^3+3pq^2},\frac{3pq^2}{p^3+3pq^2}\right)
  +\frac{3pq^2}{p^3+3pq^2}H(\frac{1}{3},\frac{1}{3},\frac{1}{3})\right\}
  \nonumber\\
  &=(p^3+3pq^2)H\left(\frac{p^3}{p^3+3pq^2},\frac{3pq^2}{p^3+3pq^2}\right)
    +3pq^2\log 3.\nonumber
\end{align}
$f(q,p)$ is obtained by interchanging $p$ with $q$.
We note that, since $p+q=1$,
$3pq^2\log 3+3qp^2\log 3=3pq\log 3$.

From working with blocks of size~3 we get
\begin{align}
  \label{equation:3Hpq}
  3H(p,q)
  &=H(p^3+3pq^2,q^3+3qp^2)
  +(p^3+3pq^2)H\left(\frac{p^3}{p^3+3pq^2},\frac{3pq^2}{p^3+3pq^2}\right)
  \nonumber\\
  &\hphantom{=H(p^3+3pq^2,}
  +(q^3+3qp^2)H\left(\frac{q^3}{q^3+3qp^2},\frac{3qp^2}{q^3+3qp^2}\right)
  +3pq\log 3.
\end{align}
For blocks of size~2
formula~(\ref{equation:FirstNewIdentity})
can be put in a more compact form in terms of capacities,
namely,
\begin{equation}
  \label{equationFirstNewIdentitySize2Blks}
  2\bigl(1-H(p,q)\bigr)
  =\biggl[1-H(p^2+q^2,2pq)\biggr]
  +\left[(p^2+q^2)\left(1-H\left(\frac{p^2}{p^2+q^2},\frac{q^2}{p^2+q^2}\right)\right)\right].
\end{equation}
Using the same method we can find a formula
analogous to formulae~(\ref{equation:FirstNewIdentity}), (\ref{equation:3Hpq})
for obtaining $nH(p,q)$ as a linear combination
of terms of the form
$H(u,v)$ where $u,v$ involve terms in
$p^n,p^{n-2}q^2,\ldots,q^n,q^{n-2}p^2,\ldots$
plus extra terms such as
$3pq\log 3$ as in
formula~(\ref{equation:3Hpq}).

%
%
%
\section{Generalizations, an Addition Formula}
\label{section:GeneralizationsAndAdditionFormula}
The result of Theorem~\ref{theorem:XYInputOutput}
can be extended to the more general case where
we take the product of $n$ binary symmetric channels
even if the channel matrices can be different
corresponding to differing $p$-values.

As an example, suppose we use the product of
2 binary symmetric channels with channel matrices
$$
  \left(
  \begin{array}{cc}
  p_1&q_1\\
  q_1&p_1\end{array}\right),
  \quad
  \left(
  \begin{array}{cc}
  p_2&q_2\\
  q_2&p_2\end{array}\right)
  .$$
Then the argument in Section~\ref{section:NewIdentities}
goes through.  To avoid being overwhelmed
by symbols we made a provisional notation change.

\begin{notationN}
  We denote by $h(p)$ the quantity
  $H(p,q)=p\log\frac{1}{p}+q\log\frac{1}{q}$.
\end{notationN}

Then we arrive at the following addition formula
\begin{align}
h(p_1)+&h(p_2)\nonumber\\
    &=h(p_1p_2+q_1q_2)
      +(p_1p_2+q_1q_2)\,h\left(\frac{p_1p_2}{p_1p_2+q_1q_2}\right)\nonumber\\
      &\qquad +(p_1q_2+p_2q_1)\,h\left(\frac{p_1q_2}{p_1q_2+p_2q_1}\right).
\end{align}
Similarly to the above
we can derive a formula for
$h(p_1)+h(p_2)+\cdots+h(p_n)$.

%
%
%
\section{The Shannon Limit and Applications to Cryptography}
\label{section:ShannonLimitAppnsCrypt}
The above method of using blocks of various sizes
is reminiscent of the algorithm for the key exchange in
\cite{BruenForcinitoWehlau}
which relates to earlier work in~\cite{Bennett}, \cite{MaurerWolf}
and others.
Indeed the identities above were informed by the
details of the algorithm.

The algorithm starts with two arrays
$(a_1,\ldots,a_n)$ and $(b_1,\ldots,b_n)$.
We assume that the set of events
$\{a_i=b_i\}$ is an independent set with
$p=Pr(a_i=b_i)$.
We subdivide $A,B$ into corresponding sub-blocks of size~$t$,
where $t$ divides $n$.
Exchanging parities by public discussion we end up with
new shorter sub-arrays $A_1,B_1$, where the probabilities
of corresponding entries being equal are independent
with probability $p_1>p$.
Eventually after $m$ iterations we end up
with a common secret key $A_m=B_m$.

Let us take an example.
Start with two binary arrays
$(a_1,\ldots,a_n)$ and $(b_1,\ldots,b_n)$
of length $n$ with $n$ even, $n=2t$, say.
We subdivide
the arrays into corresponding blocks of size~2.
If the two blocks are
$\left(\begin{array}{c}a_1\\a_2\end{array}\right)$
and
$\left(\begin{array}{c}b_1\\b_2\end{array}\right)$
we discard those blocks if the parities disagree.
If the parities agree, which happens with
probability $p^2+q^2$,
we keep $a_1$ and $b_1$,
discarding $a_2$ and $b_2$.
Thus, on average, we keep
$(p^2+q^2)\frac{n}{2}$ partial blocks
and discard $\bigl[1-(p^2+q^2)\bigr]\frac{n}{2}$ blocks of size~2.

Let us suppose $n=100$ and $p=0.7$.
From Theorem~\ref{theorem:XYInputOutput} part~(d),
the information that $Y$ has about $X$, i.e.,
that $B$ has about $A$ is
$100[1-H(0.7,0.3)]\approx100(1-0.8813)\approx 11.87$~Shannon bits.

We are seeking to find a sub-array of $A,B$ such that
corresponding bits are equal.
Our method is to publicly exchange parities.
The length of this secret key
\index{key length}
will be at most~11.

Back to the algorithm.
$A,B$ keep on average $(50)(p^2+q^2)$ blocks of size~2,
i.e., $(50)(0.58)=29$ blocks of size~2.
$A$ and $B$ remove the bottom element of each block.
We are left with 29 pairs of elements $(a_1,b_1)$.
The probability that $a_1=b_1$ given that
$a_1+a_2=b_1+b_2$ is
$\frac{p^2}{p^2+q^2}$,
i.e.,
$\frac{(0.7)^2}{(0.7)^2+(0.3)^2}=\frac{0.49}{0.58}\approx 0.845$.
Next, $1-H(0.845,0.155)\approx(1-0.6221)\approx 0.3779$.
To summarize, we started with
100 pairs $(a_i,b_i)$ with
$Pr(a_i=b_i)=0.7$.
The information revealed to $B$ by $A$ is
{\em $(100)(1-H(0.7,0.3))=11.87$ Shannon bits of information}.
After the first step of the algorithm
we are left with 29 pairs $(a_j,b_j)$ with
$Pr(a_j=b_j)=0.845$.
The  amount of information revealed to the remnant of $B$
by the remnant of $A$ is
{\em $29(0.3779)\approx 10.96$ Shannon bits of information}.
So we have ``wasted'' less than 1 bit,
i.e. the wastage is about 8\%.
Mathematically we have
$100\bigl(1-H(0.7,0.3)\bigr)=29\bigl(1-H(0.845,0.155)\bigr)+$~Wastage.
In general we have
$$n[1-H(p,q)]=
\frac{n}{2}(p^2+q^2)\left[1-H\left(\frac{p^2}{p^2+q^2},\frac{q^2}{p^2+q^2}\right)\right]+W,$$
where $W$ denotes the wastage.
Dividing by $\frac{n}{2}$ we get
$$2[1-H(p,q)]=
(p^2+q^2)\left[1-H\left(\frac{p^2}{p^2+q^2},\frac{q^2}{p^2+q^2}\right)\right]
+\frac{2W}{n}.$$

\noindent
{\em Comparing with formula(\ref{equationFirstNewIdentitySize2Blks})}
we see that
$W=\frac{n}{2}\bigl[1-H(p^2+q^2,2pq)\bigr]$.
In this case $W=50\bigl[1-H(0.58,0.42)\bigr]\approx 50(1-0.9815)
=(50)(0.0185)=0.925$.

{\em To sum up then
the new identities tell us exactly how much information
was wasted and not utilized.}
They also tell us, in conjunction with the algorithm,
the optimum size of the sub-blocks at each stage.

One of the original motivations for work in coding theory
was that
the Shannon fundamental theorem showed
how capacity was the bound for accurate communication
but the problem was to construct linear codes
or other codes such as turbo codes
that came close to the bound.

Here we have an analogous situation.
In the example just considered
the maximum length of a
\index{key length}
cryptographic common secret key
\index{cryptography!secret key@\textsl{secret key}}
obtained as a common subset of $A,B$
is bounded by
$n(1-H(p))$.
The problem is to find algorithms which produce such a
common secret key coming close to this Shannon bound
of $n\bigl(1-H(p)\bigr)$.

This work nicely illustrates
the inter-connections between 
codes, cryptography, and information theory.
Information theory tells us the bound.
The identities tell us the size of the sub-blocks
for constructing 
a common secret key
which attains, or gets close to,
the information theory bound.

Coding theory is then used to ensure that the
two communicating parties have a common secret key
by using the hash function
described in the algorithm using
a code~$C$.
Error correction can ensure that
the common secret key
can be obtained without using
another round of the algorithm
(thereby shortening the common
key) if the difference between
the keys of $A$ and $B$ 
is less than the minimum distance
of the dual code of $C$.
This improves on the standard method of
checking parities of random subsets
of the keys $A_m,B_m$ at the last stage.

%
%
%
%
\section{Concluding Remarks.}
\begin{enumerate}
\item
  Please see 
  Chapter~25 of 
  the forthcoming
  {\em Cryptography, Information Theory, and Error-Correction:
    A Handbook for the Twenty-First Century},
  by  Bruen, Forcinito, and McQuillan, \cite{BruenForcinitoMcQuillan},
  for background information, additional details, 
  and related material.

\item
  Standard tables for entropy list values to two decimal places.
  When $p$ is close to 1 interpolation
  for three decimal places is difficult as
  $h(p)$ is very steeply sloped near $p=1$.
  Formula~\ref{equation:FirstNewIdentity} 
  may help since $\frac{p^2}{p^2+q^2}$ is less than $p$,
  and the formula can be re-iterated.

\item
  In~\cite{BruenForcinitoWehlau}
  the emphasis is on the situation where the eavesdropper
  has no initial information.
  The case where the eavesdropper has initial
  information is discussed in~\cite{MaurerWolf}.
  In Section 7 of~\cite{BruenForcinitoWehlau}
  the quoted result uses Renyi entropy 
  rather than Shannon entropy.

\item
  The methods in this note suggest possible generalizations
  which we do not pursue here.

\end{enumerate}

\bigskip\noindent
{\bf
Acknowledgement:}
The author thanks Drs Mario Forcinito, James McQuillan
and David Wehlau for their help and encouragement with 
this work.

\bigskip
\bibliography{BruenNewIdentities}

\end{document}